# Event Display in the JUNO Experiment

**Jiang Zhu[a,1], Zhengyun You[a,2*] and Yumei Zhang[b,3],**
**on behalf of the JUNO Collaboration**

[a] School of Physics, Sun Yat-sen University, Guangzhou 510275, China
[b] Sino-French Institute of Nuclear Engineering and Technology, Sun Yat-sen University, Guangzhou 510275, China

Email: [1]zhuj38@mail2.sysu.edu.cn, [2]youzhy5@mail.sysu.edu.cn,
[3]zhangym26@mail.sysu.edu.cn

**Abstract**. The current event display system in the offline software of Jiangmen Underground Neutrino Observatory Experiment(JUNO) is based on the ROOT EVE package. We use Unity, a renowned game engine, to improve its performance and make it available on different platforms. Compared to ROOT, Unity provides a more vivid demonstration for high energy physics experiments and can be ported to different platforms easily. We build a tool for event display in JUNO with Unity. It provides us an intuitive way to observe the detector model, the particle trajectories and the hit distributions.

## 1. Introduction

An event display visualizes a high energy physics experiment, showing the detector structure and event information. It can also be used to improve the reconstruction algorithm and help physics analyses. With the event display system, physicists can understand the detector structure quickly and can observe some special events intuitively.

Jiangmen Underground Neutrino Observatory Experiment (JUNO) is a next-generation reactor neutrino experiment. It is located in Jiangmen City, Guangdong Province, China, which is in a distance of 53km from both of Yangjiang and Taishan nuclear power plants [1]. The main scientific purpose of JUNO is to measure the mass hierarchy of the three generations of neutrinos using inverse beta decay [2]. The central detector of JUNO is a 35.4m-diameter acrylic sphere with 20000 tons liquid scintillator inside and more than 17000 Photo Multiple Tubes(PMTs) around the sphere's surface. The whole detector is surrounded by water and there is a muon tracker on the top of it [1].

Based on the demands of the JUNO experiment, the collaboration develops the offline software system for JUNO based on the software framework SNiPER [3], (Software for Non-collider Physics ExpeRiment). The current event display system of JUNO is integrated in the JUNO offline software system and it is based on ROOT [4]. We use the Event Visualization Environment(EVE) package [5] to build this application for JUNO to show the detector structure, to animate the event hits, to visualize the reconstruction vertex and other event information. At the same time, we are also developing a new event display system based on the Unity engine to make the event display system less dependent on the JUNO offline software system so that the users can run the event display on their own PCs.

∗ Supported by National Natural Science Foundation of China (11405279, 11675275) and the Strategic Priority Research Program of Chinese Academy of Sciences (XDA10010900).

## 2. Event display based on ROOT

ROOT is a data analysis framework for high energy physics [4], based on C++. The EVE package is a visualization library in ROOT. It can manage the geometry objects and GUI components at a high level and render the image with OpenGL [5].

In the framework of ROOT, we use the EVE package to build the current event display system SERENA, which stands for Software for Event display with ROOT EVE in Neutrino Analysis. Figure 1 shows the current scheme of the event display in JUNO offline software.

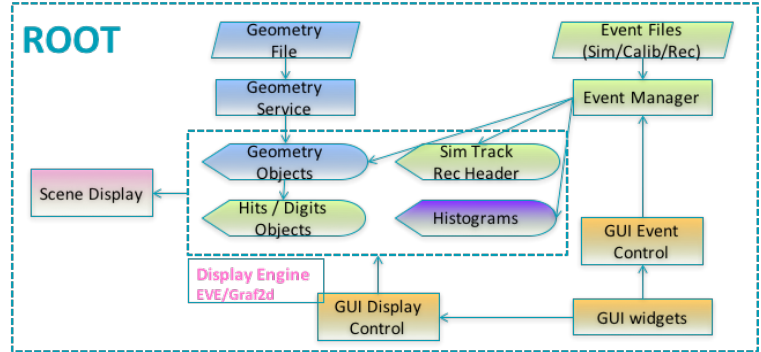

**Figure 1.** The current scheme of the event display in JUNO offline software

SERENA loads the geometry file of the JUNO detector to get the structure information and uses the EVE package to generate the visual objects. Then it reads the event information from the simulation/calibration/reconstruction files generated by the JUNO offline software. After getting the PMT hits information in each event, it changes the visual effects of the objects generated with the geometry file, builds the reconstruction vertex and shows the detailed information of the event. Users are able to select the event that they want to display with the GUI interface. In addition, it can read the data from the event file directly and project the hit distributions into azimuth angle to get the 2D plot so that users can observe the distribution in 2D projection view.

Figure 2 shows the GUI of SERENA. In the middle is the display windows, showing the PMT hit distributions in the central detector. The left side has the visual object selector, with which users can select the specific structure components they want to display or not, such as hiding the water pool, the acrylic ball or the veto PMTs. The right side has the widgets to control the event, where users can choose the specific event and demonstrate the PMT hit animation. With SERENA we can show the detector structure, animate the event hit, draw the construction vertex and make a comparison between simulation and reconstruction for reconstruction algorithm tuning.

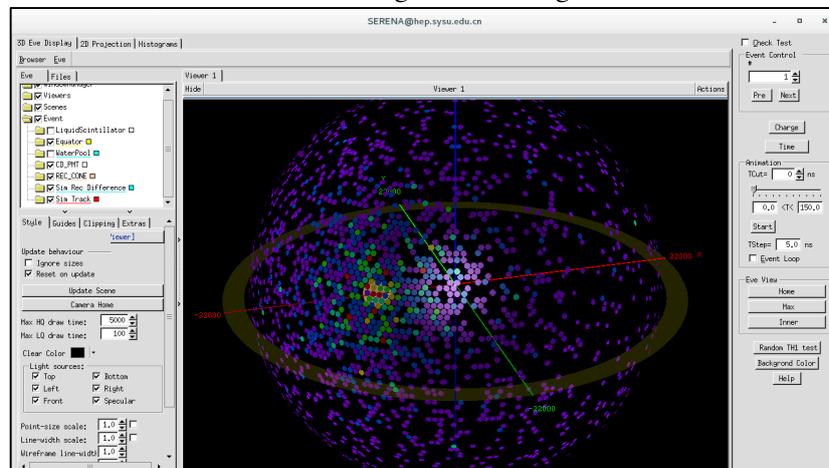

**Figure 2.** GUI of SERENA

For now, SERENA has satisfied most of the demands of JUNO experiment for event display. It is integrated in the JUNO offline software system as a package. Usually the JUNO offline software will be set up on servers with Scientific Linux system. If users want to run the display system on their own laptops, they will have to install some plugins like XQuartz or VNC to display the screen locally. The situation is even more complex as different users use different operation systems. An alternative event display system, which is less dependent on the JUNO offline software, might be helpful for users. We consider using the Unity engine as suitable for building such an independent event display system.

**3. Event display based on Unity**

*3.1. Introduction to Unity*
Unity [6] is a renowned game engine. Many interesting games like the Firewatch, Monument Valley, and Disney Crossy Road is developed with the Unity engine. Unity is known for its support for multiple platforms, which means an application developed with Unity can be built for other platforms very easily. The developers do not need to spend too much time on porting their application to make it available on Windows, Linux, macOS or in the browser. Not only famous game companies, but also many independent developers like using Unity to build their games. The personal edition of Unity is free for people who do not make sizable profits with it and the personal edition includes most of the functions of Unity. What's more, Unity is not just for games and there are many successful projects in education, simulation and visualization. In the field of high energy physics, there is a project developed by CERN called CAMELIA (Cross-platform Atlas Multimedia Educational Lab for Interactive Analysis) [7]. It can be used to visualize the detector of ATLAS experiment and analyze the collider events in LHC. When developing with Unity, developers script with C# or JavaScript. There are many useful modules for developers to build their GUI very quickly, like button, slider, toggle and so on.

*3.2. Requirements analysis*
In contrast to the event display system based on ROOT, the new event display system based on Unity is built as a client so that the system can be less dependent on the JUNO offline software and can run on users' own PCs only after importing the event data. As an event display system in high energy physics experiment, it should do the visualization of the detector structure, restore the process of event hits, load reconstruction data and make comparisons between reconstruction and simulation. At the same time, we try to build this system as a client to make it easy to upgrade into the online event display system for further development.

*3.3. Scheme and data flow*
The new event display system based on Unity has almost the same data flow as the one based on ROOT, as figure 3 shows. In JUNO offline software, the original detector description file is in text format. Then the offline software builds the detector structure with Geant4 and runs the simulation process. After that, it can output the geometry file in GDML format or store it in a ROOT file as ROOT geometry objects, together with the simulation data. In the new event display system, we also get the detector structure from the geometry file, read the event data from the event file generated by JUNO offline and obtain the reconstruction vertex and energy from the reconstruction file. All the data files generated by JUNO offline are in ROOT format, including the geometry, simulation, calibration and reconstruction files. Without the ROOT framework and the event data model of JUNO offline [8], the data files cannot be read by the new event display system. It is inefficient to add the whole ROOT library and JUNO data model into the new application as it would consume disk space and computing resources. It also should be noted that JUNO offline is still in the development stage and the data model may be adjusted in the coming versions. Instead, we write a ROOT macro to output the data we need into text format and then the new display system can read the text files to get the event data and detector structure. The macro will be available for users in JUNO offline. It is more efficient that we do the data conversion in the computing node, where we set up the JUNO offline software.

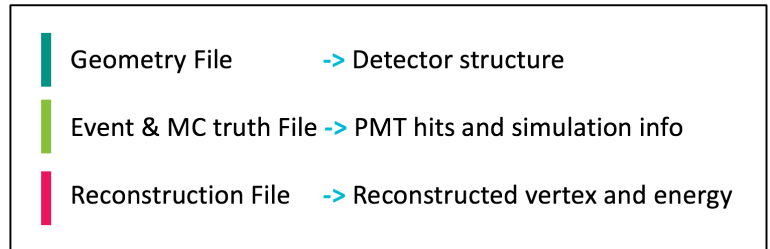

**Figure 3.** Data flow of the event display based on Unity

*3.4. Interface and functions*

With the data flow in figure 3, we realize three main functions in the new event display system, including the visualization of detector structure, animation of event hits and comparison between simulation and reconstruction. Figure 4 is the visualization of the JUNO detector. From outside to inside, the cylinder on the outermost layer is the water pool that contains the central detector and there are about 1700 veto PMTs in the water pool. The next layer is the central detector PMTs around the surface of the acrylic sphere (only half is shown in figure 4). And the inner layer is the liquid scintillator.

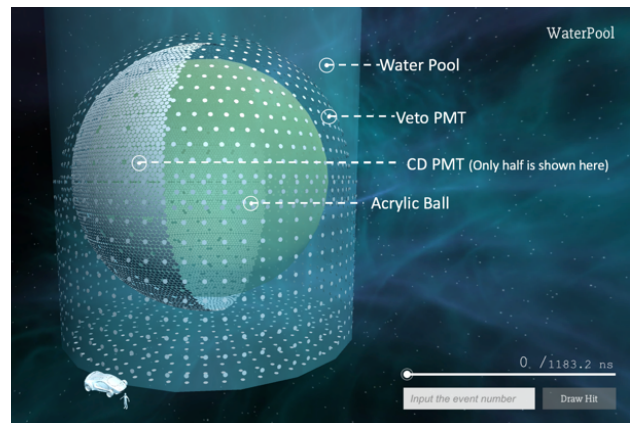

**Figure 4.** Visualization of JUNO detector

Figure 5 shows a PMT hit distribution in the detector. Users can input the event number they want to display in the lower right corner of the window, and then click the draw button to trigger the animation of the central detector PMTs. The circles that appear in figure 5 represent the PMTs that have already been hit and the color of the circles stands for the number of photons they receive. Users are also able to freeze time to observe the specific distribution at that moment by dragging the slider in the lower right corner. Furthermore, the system provides detailed information about detector components if users move their mouse cursor onto a specific component. Figure 5 shows the identifier, the coordinates and the first hit time of the PMT.

It also provides visualization of the reconstruction vertex and energy. Figure 6 shows the initial point, the energy deposition point and the reconstruction point. Users can observe the difference between simulation and reconstruction intuitively and use it for reconstruction algorithm tuning.

Different to the ROOT based event display, to maximize the view of users, we hide the detailed settings from the main interface. A shortcut is set up to call out the option menu of this system, where users can select the objects to display, change the lighting effect and view more information about the event file. In addition, there is also a camera system for users to observe the detector and events from different angles as they want.

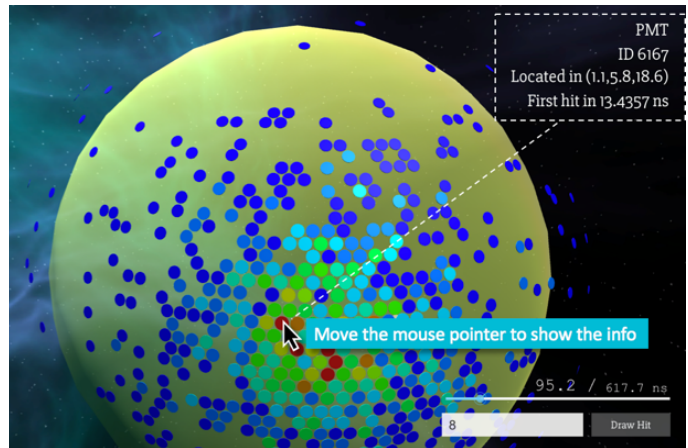

**Figure 5.** PMT hit distribution

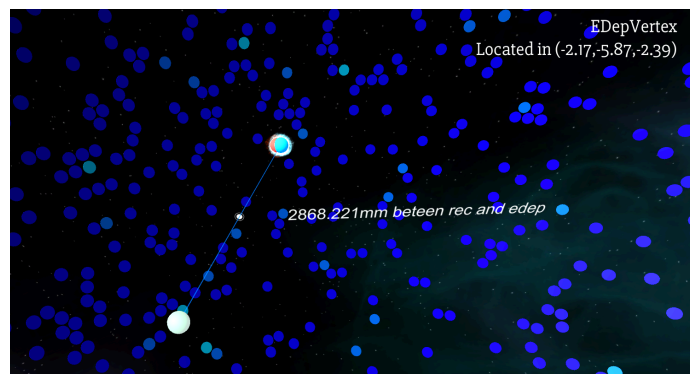

**Figure 6.** Comparison of vertexes

*3.5. Test of portability*
We have built the new event display system on different platforms to test the portability of the application developed by Unity and to see if the software can run successfully in the corresponding operating systems. Considering the operation system that the members of JUNO usually use, we have built the event display in Windows 8.1, Scientific Linux 7.2 and macOS 10.12. The application is available on these platforms. What's more, we also build it into html format and use Chrome 58.0 to run it with a lower display resolution, so it's possible for us to build an online event display system with Unity.

**4. Comparison of the two event display systems**
Currently there are two event display systems for JUNO. The first one, SERENA, is in the framework of ROOT and the second one is based on the Unity engine.

The event display system based on ROOT is approaching full functionality. It has realized the visualization of the JUNO detector and users can use it for reconstruction algorithm tuning and physic analysis. Compared to the Unity one, it has the advantage that it can easily read the data in ROOT format since it is in the framework of ROOT. On the other hand, SERENA is integrated in the JUNO offline software so users need to set up the whole offline software to run the event display. If the users want to receive the screen of event display remotely, they need to install some other plugins. And the visual effects of SERENA are also limited by the ROOT EVE package itself.

The biggest advantage of the event display based on Unity is that it can be deployed onto different platforms quickly after development. And it is built as a client so users can run the event display on their own PCs without setting up the whole offline software. Besides, as a game engine, it is easier for the developer to realize fancier visual effects. On the contrary, the shortcoming of the system is obvious. The extra data conversion is necessary to get the event data from the ROOT format file generated by the offline software.

## 5. Conclusion

Considering the demands of the JUNO experiment, we have built two event display systems to meet the requirements of users. The original event display based on ROOT, SERENA, is approaching full functionality and the new event display system based on Unity is essentially available. With the Unity engine, we can easily realize better visualization effects. The new event display system is highly portable and it is possible to be upgraded into the online event display system.

*The authors would like to thank the members of JUNO for their valuable discussions and suggestions, and thank the support from National Natural Science Foundation of China (11405279, 11675275) and the Strategic Priority Research Program of Chinese Academy of Sciences (XDA10010900).*

**References**
[1]  Adam T, An F, An G, et al. JUNO conceptual design report. *arXiv preprint arXiv:1508.07166*.
[2]  An F, An G, An Q, Antonelli V, Baussan E, Beacom J et al. Neutrino physics with JUNO. *Journal of Physics G: Nuclear and Particle Physics*. 2016;**43**(3):030401.
[3]  Zou J, Huang X, Li W, Lin T, Li T, Zhang K et al. SNiPER: an offline software framework for non-collider physics experiments. *Journal of Physics: Conference Series*. 2015;**664**(7):072053.
[4]  https://root.cern.ch
[5]  Tadel M. Overview of EVE – the event visualization environment of ROOT. *Journal of Physics: Conference Series*. 2010;**219**(4):042055.
[6]  https://unity3d.com
[7]  http://medialab.web.cern.ch/content/camelia
[8]  Li T, Xia X, Huang X, Zou J, Li W, Lin T et al. Design and development of JUNO event data model. *Chinese Physics C*. 2017;**41**(6):066201.